\documentclass[conference]{IEEEtran}
\IEEEoverridecommandlockouts
\usepackage{bbm}
\usepackage{cite}
\usepackage{hyperref}       
\usepackage{amsmath,amssymb,amsfonts}
\usepackage{graphicx}
\usepackage{textcomp}
\usepackage{xcolor}
\usepackage{float}
\usepackage{mathtools}
\usepackage{color}
\usepackage{bm}
\usepackage{booktabs}
\usepackage[noend]{algpseudocode}
\usepackage{algorithm,algcompatible,tabularx}
\usepackage{tikz}
\usepackage{balance}
\usepackage[utf8]{inputenc}
\usepackage[T1]{fontenc} 
\usepackage{multicol}
\usepackage{balance}
\usepackage{caption}
\usepackage{algorithm}
\usepackage{enumitem}
\usepackage{mwe}
\usepackage{diagbox}
\usepackage{multirow}
\usepackage{makecell}
\usepackage{kantlipsum}
\usepackage{svg}
\usepackage[noend]{algpseudocode}

\newcommand{\x}{{\mathbf x}}

\newcommand{\y}{{\bf y}}

\DeclareMathOperator*{\argmax}{arg\,max}

\makeatletter
\newcommand{\multiline}[1]{%
  \begin{tabularx}{\dimexpr\linewidth-\ALG@thistlm}[t]{@{}X@{}}
    #1
  \end{tabularx}
}
\makeatother

\title{Particle Filtering Under General Regime Switching \\
\thanks{This work was carried out thanks to the support of the National Science Foundation (NSF) under Awards CCF-1617986 and CCF-1618999. The authors thank the Research Computing and Cyberinfrastructure and the Institute for Advanced Computational Science at Stony Brook University for access to the high-performance SeaWulf computing system sponsored by NSF (\# 1531492).}}

\author{\IEEEauthorblockN{Yousef El-Laham, Liu Yang, Petar M. Djuri\'{c}, M\'{o}nica F. Bugallo}
\IEEEauthorblockA{Department of Electrical \& Computer Engineering \\
Stony Brook University, Stony Brook (USA)} \vspace{-0.45cm}  \\
{\{yousef.ellaham, liu.yang.2, petar.djuric, monica.bugallo\}@stonybrook.edu}}

\begin{document}
\maketitle
\begin{abstract}
In this paper, we consider a new framework for particle filtering under model uncertainty that operates beyond the scope of Markovian switching systems. Specifically, we develop a novel particle filtering algorithm that applies to general regime switching systems,  where the model index is augmented as an unknown time-varying parameter in the system. The proposed approach does not require the use of multiple filters and can maintain a diverse set of particles for each considered model through appropriate choice of the particle filtering proposal distribution. The flexibility of the proposed approach allows for long-term dependencies between the models, which enables its use to a wider variety of real-world applications. We validate the method on a synthetic data experiment and show that it outperforms state-of-the-art multiple model particle filtering approaches that require the use of multiple filters. 
\end{abstract}

\section{Introduction}
\label{sec: introduction}
In the past three decades, particle filtering (PF) \cite{gordon1993novel,djuric2003particle} has emerged as one of the most powerful statistical tools for online state estimation in {dynamical} systems. PF methods approximate the posterior distribution of an unknown time-varying parameter vector in a state-space model (SSM) using a set of weighted samples. The samples, which are also called particles, are drawn from a probability distribution called the \emph{proposal distribution} and are weighted properly according to the principle of importance sampling \cite{robert2013monte}. Unlike Kalman filtering and its extensions \cite{kalman1960new,julier2004unscented}, PF can deal with SSMs that exhibit both nonlinearities and non-Gaussianities. This flexibility has allowed PF methods to thrive in many applications in fields as diverse as signal processing, economics, neuroscience, epidemiology, and ecology \cite{karlsson2000auxiliary, creal2012survey, brockwell2004recursive, jegat2008early, dowd2006sequential}.

Model uncertainty introduces an additional layer of complexity to stochastic filtering that is generally difficult to deal with. In this situation, one must determine the model that best represents the system of interest from a set of candidate models, while also jointly estimating the unknown time-varying parameters of the chosen model. The issue of model uncertainty is further complicated if the model can switch from one time instant to the next. In signal processing, the well-known problem of tracking a maneuvering target falls within this class of model selection problems \cite{bugallo2007performance}. The trajectory of a maneuvering target is represented via a Markovian switching system (i.e., jump Markov systems) \cite{liang2009multiple}, where the model dynamics change according to the state of a discrete-time, discrete-state Markov chain. More generally, however, systems whose models (or regimes) can change from one time instant to the next are referred to as \emph{regime-switching systems}. 

There have been mainly two classes of solutions proposed in the PF literature to deal with the challenge of model uncertainty. In the first class of solutions, a model index that references the candidate models is augmented as an unknown state in the system and is jointly estimated with the unknown time-varying parameters using a single particle filter \cite{liang2009multiple, mcginnity2000multiple}. While this solution is simple to implement and straightforwardly tackles the joint estimation problem, the disadvantage is that the number of samples assigned to each candidate model cannot be controlled, which can lead to numerical issues and a lack of diversity in the considered models. The second class of solutions employs the use of a bank of particle filters that operate in parallel. Each filter is conditioned on one of the candidate models and state estimates are obtained by weighting the results of each of the filters and then fusing them. In \cite{liu2011instantaneous}, the filters are weighted according to the posterior probability of their respective models, while in \cite{urteaga2016sequential}, they are weighted according to the predictive powers of their respective models. Hybrid solutions which combine this class of approaches with interacting {multiple} models (IMM) filter have also been proposed to deal with Markovian switching systems \cite{boers2003interacting}.  Unfortunately, because a separate filter is required for each model, this class of approaches can be computationally intensive if the number of candidate models is large. We remark that, to the best of our knowledge, non-heuristic implementations of both methods have not been formulated for more general systems, which may exhibit long-term dependencies in the regime switching dynamics beyond those of the Markovian switching systems.   

In this work, we propose a novel PF algorithm for general regime switching systems. Similar to the aforementioned solutions, the proposed PF method augments the model index as an unknown in the system that is jointly estimated with the time-varying parameters. Unlike previous approaches in the literature, a diverse set of candidate models can always be considered in the proposed algorithm through appropriate choice of the model index proposal distribution. Furthermore, since our method is not restricted to Markovian switching systems, it can handle more complicated processes that may describe the regime switching dynamics, such as the P\'{o}lya urn process \cite{gouet1997strong}. Simulation results for synthetic data experiments validate the performance of the proposed approach. 

\section{Problem Formulation}
\label{sec: problem_formualtion}
\begin{figure}[t]
    \centering
    \includegraphics[trim={0cm 7cm 16cm 0cm},clip, width=0.75\linewidth]{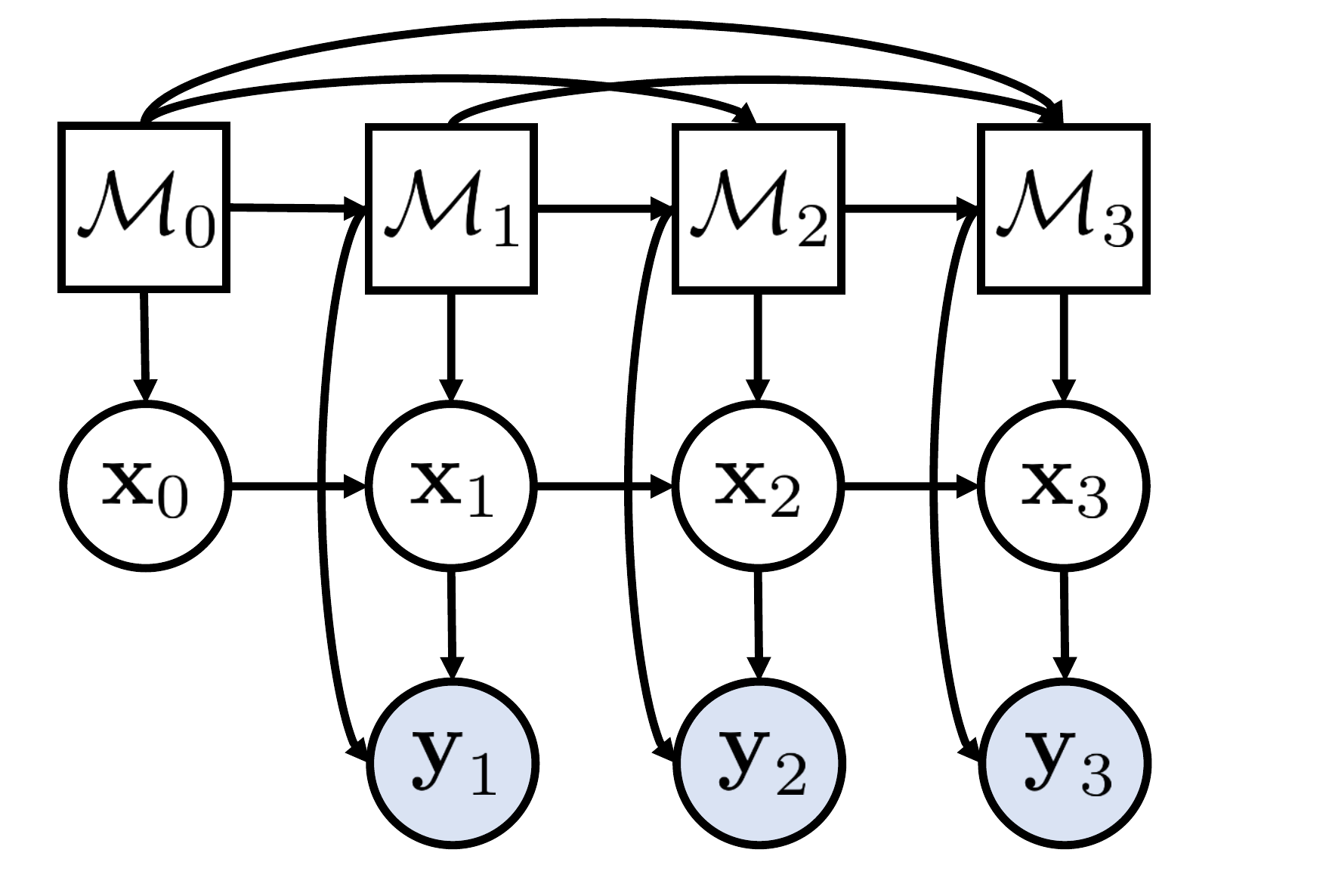}
    \caption{The considered regime switching SSM formulation for a time horizon $T=3$. The probability of a model depends on the complete history of models at previous time instants.}
    \label{fig: graphical_model}
\end{figure}

Let $\x_t\in\mathbb{R}^{d_x}$ denote a latent state vector, $\y_t\in\mathbb{R}^{d_y}$ denote a measurement vector, and $\mathcal{M}_t\in\{1,\ldots,K\}$ denote a model index from a set of $K$ candidate models, where $t$ denotes time index. We consider the generic stochastic filtering problem under model uncertainty over a fixed time horizon $T$. The generative process is assumed to have the following form:
\begin{align}
    \mathcal{M}_t &\sim p(\mathcal{M}_t|\mathcal{M}_{0:t-1}), \\
    \x_t &\sim p(\x_t|\x_{t-1}, \mathcal{M}_t), \\
    \y_t &\sim p(\y_t|\x_t, \mathcal{M}_t),
\end{align}
for $t=1,\ldots, T$, where the initial model $\mathcal{M}_0$ is distributed according to $\mathcal{M}_0\sim p(\mathcal{M}_0)$ and the initial latent state $\x_0$ is distributed according to $\x_0\sim p(\x_0|\mathcal{M}_0)$. A graphical representation of the system is shown in Fig. \ref{fig: graphical_model} for $T=3$.

Our goal is to jointly infer the unknown states $\x_{0:T}\triangleq[\x_0,\ldots,\x_T]\in\mathbb{R}^{d_x\times (T+1)}$ and the sequence of unknown models $\mathcal{M}_{0:T}\triangleq (\mathcal{M}_0,\ldots,\mathcal{M}_T)$ based on the given observations $\y_{1:T}\triangleq [\y_1,\ldots,\y_T]\in\mathbb{R}^{d_y\times T}$ under the Bayesian paradigm. In other words, we would like to approximate the joint posterior distribution $p(\x_{0:T},\mathcal{M}_{0:T}|\y_{1:T})$ in a recursive manner using a Bayesian filtering solution.  


\section{Regime Switching Particle Filtering}
\label{sec: proposed_work}
In this section, we derive a generalized regime switching PF (RSPF) algorithm. We begin by first establishing the recursiveness of the joint distribution and then we derive the importance weights of the particles in the novel PF algorithm. We discuss different strategies for sampling models which allow for model diversity. Finally, we elaborate on how one can use the proposed algorithm to obtain the maximum \emph{a posteriori} (MAP) estimate of the model at each time instant. We summarize the proposed approach in Algorithm \ref{alg: generic_RSPF}.

\begin{algorithm}[t]
\caption{Regime Switching Particle Filtering (RSPF)}
\label{alg: generic_RSPF}
  \begin{algorithmic}[1]
    \STATE \textbf{Initialization:} Draw $N$ samples from the prior of the initial model to determine the model indexes
    \begin{equation*}
        \mathcal{M}_0^{(n)}\sim p(\mathcal{M}_0), \quad n=1,\ldots,N,
    \end{equation*} 
    and draw $N$ samples from the prior of the initial state conditioned on the sampled model indexes 
    \begin{equation*}
        \x_0^{(n)}\sim p(\x_0|\mathcal{M}_0^{(n)}), \quad n=1,\ldots,N.
    \end{equation*} 
    Set the weights as $\tilde w_0^{(n)}=\frac{1}{N}$ for $n=1,\ldots, N$.
    \FOR{$t=1,\ldots,T$}
    \STATE \multiline{\textbf{Sampling models:} Draw $N$ samples from the model index proposal distribution}
    \begin{equation*}
        \mathcal{M}_t^{(n)}\sim q(\mathcal{M}_t|\mathcal{M}_{0:t-1}^{(n)}), \quad n=1,\ldots,N.
    \end{equation*}
    \STATE \multiline{\textbf{Sampling states:} Draw $N$ samples of the states conditioned on the drawn models}
    \begin{equation*}
        \x_t^{(n)}\sim q(\x_t|\x_{t-1}^{(n)}, \mathcal{M}_t^{(n)}, \y_t), \quad n=1,\ldots,N.
    \end{equation*}
    \STATE \multiline{\textbf{Weighting:} Compute the weights $\{\tilde w_t^{(n)}\}_{n=1}^N$ according to \eqref{eq: importance_weights_regime_switching} and normalize them as}
    \begin{equation*}
        w_t^{(n)} = \frac{\tilde w_t^{(n)}}{\sum_{j=1}^N \tilde w_t^{(j)}}, \quad n=1,\ldots,N.
    \end{equation*}
    \STATE \multiline{\textbf{Model selection:} Determine the model index estimate $\hat{ \mathcal{M}}_t$ by solving the following maximization problem
    \begin{equation*}
        \hat{\mathcal{M}}_t = \argmax_{k\in\{1,\ldots,K\}} p(\mathcal{M}_t=k|\y_{1:t})
    \end{equation*}
    where \resizebox{0.7\hsize}{!}{$p(\mathcal{M}_t=k|\y_{1:t})\approx \sum_{n=1}^N  w_t^{(n)}\mathbbm{1}(\mathcal{M}_t^{(n)}=k)$} for all $k$.}
    \STATE \multiline{\textbf{State estimation:} Obtain the state estimate as}
    \begin{equation*}
        \hat\x_t = \sum_{n=1}^N w_t^{(n)}\x_t^{(n)}.
    \end{equation*}
    \STATE \multiline{\textbf{Resampling:} If necessary, resample the model indexes and the states using multinomial resampling and set the weights as $\tilde w_t^{(n)}= \frac{1}{N}$ for $n=1,\ldots,N$.}
    \ENDFOR
  \end{algorithmic}
\end{algorithm}

\subsection{Deriving the Joint Distribution}
At time $t$, the distribution of interest is $p(\x_{0:t},\mathcal{M}_{0:t}|\y_{1:t})$. This joint distribution can be decomposed as
\begin{equation}
    p(\x_{0:t},\mathcal{M}_{0:t}|\y_{1:t})=p(\x_{0:t}| \y_{1:t}, \mathcal{M}_{0:t})p(\mathcal{M}_{0:t}|\y_{1:t}),
\end{equation}
where $p(\x_{0:t}|\y_{1:t},\mathcal{M}_{0:t})$ is the posterior of the state trajectory $\x_{0:t}$ conditioned on the model sequence $\mathcal{M}_{0:t}$ and $p(\mathcal{M}_{0:t}|\y_{1:t})$ is the marginal posterior distribution of the model sequence. Following previous analysis in the Bayesian filtering literature, we can readily deduce the conditional posterior of the state trajectory $p(\x_{0:t}|\y_{1:t},\mathcal{M}_{0:t})$ as
\begin{equation}
\begin{split}
&p(\x_{0:t}|\y_{1:t},\mathcal{M}_{0:t}) = \\ &\frac{p(\y_t|\x_t,\mathcal{M}_t)p(\x_t|\x_{t-1},\mathcal{M}_t)p(\x_{0:t-1}|\y_{1:t-1},\mathcal{M}_{0:t-1})}{p(\y_t|\y_{1:t-1},\mathcal{M}_{0:t})}.
\end{split}
\end{equation}
Remark that we have made the appropriate assumption that $p(\x_{0:t-1}|\y_{1:t-1},\mathcal{M}_{0:t})=p(\x_{0:t-1}|\y_{1:t-1},\mathcal{M}_{0:t-1})$, since the conditional posterior at time instant $t-1$ does not depend on the model at time $t$. Applying Bayes' theorem to the marginal posterior of the model sequence $p(\mathcal{M}_{0:t}|\y_{1:t})$, we also have
\begin{align}
    &p(\mathcal{M}_{0:t}|\y_{1:t})=\frac{p(\y_t|\y_{1:t-1},\mathcal{M}_{0:t})p(\mathcal{M}_{0:t}|\y_{1:t-1})}{p(\y_t|\y_{1:t-1})}.
\end{align}
We decompose $p(\mathcal{M}_{0:t}|\y_{1:t-1})$ as 
\begin{align}
    \resizebox{0.91\hsize}{!}{$p(\mathcal{M}_{0:t}|\y_{1:t-1}) = p(\mathcal{M}_t|\mathcal{M}_{0:t-1},\y_{1:t-1})p(\mathcal{M}_{0:t-1}|\y_{1:t-1}),$}
\end{align}
where one can show that 
\begin{align}
    &p(\mathcal{M}_t|\mathcal{M}_{0:t-1},\y_{1:t-1})=\resizebox{0.475\hsize}{!}{$\frac{p(\y_{1:t-1}|\mathcal{M}_{0:t})p(\mathcal{M}_t|\mathcal{M}_{0:t-1})}{p(\y_{1:t-1}|\mathcal{M}_{0:t-1})}$} \\
    &=\frac{p(\mathcal{M}_t|\mathcal{M}_{0:t-1})}{p(\y_{1:t-1}|\mathcal{M}_{0:t-1})}\int p(\x_{0:t-1},\y_{1:t-1}|\mathcal{M}_{0:t}) d\x_{0:t-1} \\
    &=\frac{p(\mathcal{M}_t|\mathcal{M}_{0:t-1})}{p(\y_{1:t-1}|\mathcal{M}_{0:t-1})}p(\y_{1:t-1}|\mathcal{M}_{0:t-1}) \\
    &=p(\mathcal{M}_t|\mathcal{M}_{0:t-1}),
\end{align}
since the joint distribution of $\x_{0:t-1}$ and $\y_{1:t-1}$ is conditionally independent from model $\mathcal{M}_t$ given the sequence of models up to time instant $t-1$. We can now establish the recursive solution to the joint posterior $p(\x_{0:t},\mathcal{M}_{0:t}|\y_{1:t})$ as
\begin{equation}
    \label{eq: recursive_joint}
    \begin{split}
    p(\x_{0:t},&\mathcal{M}_{0:t}|\y_{1:t}) 
     \propto p(\y_t|\x_t,\mathcal{M}_t)p(\x_t|\x_{t-1},\mathcal{M}_t) \\
     &\times p(\mathcal{M}_t|\mathcal{M}_{0:t-1})p(\x_{0:t-1},\mathcal{M}_{0:t-1}|\y_{1:t-1}).
    \end{split}
\end{equation}
\subsection{Deriving the Particle Filtering Weights}
Suppose we draw a set of $N$ samples $\{(\x_{0:t}^{(n)}, \mathcal{M}_{0:t}^{(n)})\}_{n=1}^N$, where each sample $(\x_{0:t}^{(n)}, \mathcal{M}_{0:t}^{(n)})$ is drawn from a proposal distribution $q(\x_{0:t},\mathcal{M}_{0:t}|\y_{1:t})$ for $n=1,\ldots,N$. Then, the importance weight of each sample is determined according to 
\begin{equation}
    \label{eq: standard_importance weights}
    \tilde w_t^{(n)} = \frac{p(\x_{0:t}^{(n)},\mathcal{M}_{0:t}^{(n)}|\y_{1:t})}{q(\x_{0:t}^{(n)},\mathcal{M}_{0:t}^{(n)}|\y_{1:t})}, \quad n=1,\ldots, N. 
\end{equation}
Suppose that the proposal distribution can be factored as
\begin{equation}
\begin{split}
    q(\x_{0:t},\mathcal{M}_{0:t}|&\y_{1:t}) = q(\x_{0:t-1},\mathcal{M}_{0:t-1}|\y_{1:t-1}) \\
    &\times q(\x_{t}|\x_{t-1},\mathcal{M}_t, \y_t)q(\mathcal{M}_t|\mathcal{M}_{0:t-1}).
\end{split}
\end{equation}
Combining with results from \eqref{eq: recursive_joint}, the importance weight of each sample $(\x_{0:t}^{(n)},\mathcal{M}_{0:t}^{(n)})$ can thus be determined as 
\begin{equation}
    \label{eq: importance_weights_regime_switching}
    \resizebox{0.88\hsize}{!}{$\tilde w_t^{(n)} \propto \tilde w_{t-1}^{(n)}\frac{p(\y_t|\x_t^{(n)},\mathcal{M}_t^{(n)})p(\x_t^{(n)}|\x_{t-1}^{(n)},\mathcal{M}_t^{(n)})p(\mathcal{M}_t^{(n)}|\mathcal{M}_{0:t-1}^{(n)})}{q(\x_{t}^{(n)}|\x_{t-1}^{(n)},\mathcal{M}_t^{(n)},\y_t)q(\mathcal{M}_t^{(n)}|\mathcal{M}_{0:t-1}^{(n)})}$}.
\end{equation}
For sampling the states, the bootstrap implementation of this method would assume that the proposal distribution of the states is identical to the state transition distribution, i.e., $q(\x_{t}|\x_{t-1},\mathcal{M}_t, \y_t)=p(\x_{t}|\x_{t-1},\mathcal{M}_t)$, and that the particles are resampled after each time instant. For the bootstrap implementation, the importance weights are given by
\begin{equation}
    \label{eq: importance_weights_swithcing_bootstrap}
    \tilde w_t^{(n)}\propto \frac{p(\y_t|\x_t^{(n)},\mathcal{M}_t^{(n)})p(\mathcal{M}_t^{(n)}|\mathcal{M}_{0:t-1}^{(n)})}{q(\mathcal{M}_t^{(n)}|\mathcal{M}_{0:t-1}^{(n)})},
\end{equation}
for $n=1,\ldots,N$. The obtained solution is analogous to the weighting function in bootstrap PF, except now, we must taken into account that models can change according to $p(\mathcal{M}_t|\mathcal{M}_{0:t-1})$. We remark that $p(\mathcal{M}_t|\mathcal{M}_{0:t-1})$ determines how the model $\mathcal{M}_t$ is determined from the history of models $\mathcal{M}_{0:t-1}$ and depends on the nature of system being considered. 

\subsection{Discussion on Sampling Model Indexes}
There are a variety of choices for the model index proposal distribution $q(\mathcal{M}_t|\mathcal{M}_{0:t-1})$. The most obvious choice is the \emph{bootstrap} approach, where we use the model transition function as the proposal, i.e, we set $q(\mathcal{M}_t|\mathcal{M}_{0:t-1})=p(\mathcal{M}_t|\mathcal{M}_{0:t-1})$. Then, the importance weights simply become the joint likelihood of the sampled model indexes and states. 
Alternatively, one can use a discrete \emph{uniform} proposal distribution to sample the model indexes, i.e., $q(\mathcal{M}_t=k|\mathcal{M}_{0:t-1})=\frac{1}{K}$ for all $k$. Then, each model has an equal chance to be sampled at each step of the algorithm, and thus avoiding {the possibility of a} model diversity issue. Finally, one can \emph{deterministically} sample an equal number of particles for each model. The weights in this case would be the same as if we had sampled from the discrete uniform distribution. 

\subsection{Online Maximum A Posteriori Model Selection}
In order to select the most promising model at each time instant from the set of candidate models, we need to solve the following optimization problem:
\begin{equation}
    \label{eq: map_model_selection}
    \hat{\mathcal{M}}_t = \argmax_{k\in\{1,\ldots,K\}} p(\mathcal{M}_t=k|\y_{1:t}),
\end{equation}
where $p(\mathcal{M}_t=k|\y_{1:t})$ denotes the posterior probability of the $k$th model. This posterior probability of each model can be estimated directly using the set of particles and weights
\begin{equation}
\begin{split}
    p(\mathcal{M}_t=k|\y_{1:t})\approx \frac{1}{\sum_{n=1}^N \tilde w_t^{(n)}}\sum_{n=1}^N \tilde w_t^{(n)}\mathbbm{1}(\mathcal{M}_t^{(n)}=k),
\end{split}
\end{equation}
for $k=1,\ldots, K$, where $\mathbbm{1}(\cdot)$ denotes the indicator function. Given the estimated posterior probabilities, one can obtain an approximate solution to the optimization problem in \eqref{eq: map_model_selection}.

\section{Examples of Model Sequence Dynamics}
{Here}, we {give} examples of different regime switching dynamics {that} can easily be treated using our proposed approach. 
\subsection{Independent Regime Dynamics} 
The simplest case is when the models are generated independently from one another, i.e., the joint distribution of the models can be factored as:
\begin{equation}
    \label{eq: independent_model_dynamics}
    p(\mathcal{M}_{0:T})=\prod_{t=0}^T p(\mathcal{M}_t),
\end{equation}
where the model independence assumption implies that  $p(\mathcal{M}_t|\mathcal{M}_{0:t-1})=p(\mathcal{M}_t)$. The assumption that the models are independent {may be} unrealistic for most applications and requires to specify the prior distribution of each model. 

\subsection{Markovian Switching Dynamics} 
We also consider Markovian switching systems, where the model at each time instant only depends on the model at the previous time instant. The joint distribution of the models under this assumption is given by
\begin{equation}
    \label{eq: markovian_model_dynamics}
    p(\mathcal{M}_{0:T})=p(\mathcal{M}_0)\prod_{t=1}^T p(\mathcal{M}_t|\mathcal{M}_{t-1}),
\end{equation}
where we have that $p(\mathcal{M}_t|\mathcal{M}_{0:t-1})=p(\mathcal{M}_t|\mathcal{M}_{t-1})$. Here, the model transition distribution $p(\mathcal{M}_t|\mathcal{M}_{t-1})$ is represented by a transition probability matrix ${\bf P}$
\begin{equation}
    {\bf P} = \begin{pmatrix} p_{1,1} &\ldots & p_{1,K} \\
    \vdots &\ddots &\vdots\\
    p_{K,1} &\ldots &p_{K,K}
    \end{pmatrix},
\end{equation}
where each element $p_{i,j}\triangleq p(\mathcal{M}_t=j|\mathcal{M}_{t-1}=i)$ is defined to be the probability of transitioning from model $i$ to model $j$ and each row of the matrix ${\bf P}$ satisfies $\sum_{j=1}^K p_{i,j}=1$. 

\subsection{P\'{o}lya Urn Dynamics} 
Under a more general formulation, the model at a given time instant $t$ depends on the complete sequence of models $\mathcal{M}_{0:t-1}$. Here, since there are no independence assumptions, the joint distribution of the models is given by 
\begin{equation}
    \label{eq: generalized_model_dynamics}
    p(\mathcal{M}_{0:T})=p(\mathcal{M}_0)\prod_{t=1}^T p(\mathcal{M}_t|\mathcal{M}_{0:t-1}). 
\end{equation}
{If} the number of models is finite and \emph{a priori} known, one possibility is to consider  a P\'{o}lya urn process for the regime dynamics. For the P\'{o}lya urn process, the probability of {transitioning to a particular} model at time instant $t$ depends on how many times that model was chosen in previous time instants. Let $\alpha_{k,t}=\mathbbm{1}(\mathcal{M}_t=k)$ be variable indicating if model $k$ was visited at time $t$ for $t=1,\ldots,T$ and let $\beta_{k}\in\mathbb{N}$ be any positive integer for $k=1,\ldots,K$. Then, the probability of transitioning to model $k$ at time $t$ is given by
\begin{equation}
    p(\mathcal{M}_t=k|\mathcal{M}_{0:t-1})=\frac{\beta_k+\sum_{\tau=0}^{t-1}\alpha_{k,\tau}}{\sum_{j=1}^K(\beta_j + \sum_{\tau=0}^{t-1} \alpha_{j,\tau})}.
\end{equation}

\section{Simulations}
\label{sec: simulations}
To validate the performance of the proposed RSPF, we generated synthetic measurement sequences of time length $T=50$ based on eight candidate models, with each model being of the form
\begin{align}
     \mathcal{M}_k:  
    \begin{cases*}
    x_t = a_k x_{t-1} + c_k + u_t  \\
    y_t = b_k\sqrt{|x_t|} + d_k + v_t 
    \end{cases*},
\end{align}
where the parameter settings are{ $[a_1,...,a_8]=[-0.1, -0.3, -0.5, -0.9, 0.1, 0.3, 0.5, 0.9]$,  $[c_1,...,c_8]=[0, -2, 2, -4, 0, 2, -2, 4]$, $[b_1,...,b_8]=[a_1,...,a_8]$, and $[d_1,...,d_8]=[c_1,...,c_8]$. }
The process noise $u_t$ and observation noise $v_t$ are assumed to be i.i.d. zero-mean Gaussian with equal variances, i.e., $u_t\sim\mathcal{N}(0,\sigma_u^2)$ and $v_t\sim\mathcal{N}(0,\sigma_v^2)$ with $\sigma_u^2=\sigma_v^2=0.1$. The initial state $x_0$ was generated uniformly from -0.5 to 0.5. We tested the method on two scenarios corresponding to regime switching based on Markovian dynamics and P\'{o}lya urn dynamics, respectively.

We first ran our novel algorithm with $N=2000$ particles per iteration when the model sequence dynamics is Markovian. The transition probability matrix in simulation was
\begin{equation}
     \resizebox{0.6\hsize}{!}{${\bf P} = \begin{pmatrix} 0.80 & 0.15 & \epsilon & \cdots & \epsilon \\
    \epsilon & 0.80 & 0.15 & \cdots & \epsilon \\
    \vdots &  & \ddots & \ddots & \vdots \\
    \epsilon & \cdots &   & 0.80 & 0.15 \\
     0.15 &  \epsilon & \cdots  & \epsilon & 0.80
    \end{pmatrix}$},
\end{equation}
where we set $\epsilon=\frac{1}{120}$ so that each row of ${\bf P}$ summed to 1. 

\begin{figure}[t]
\centering
 \includegraphics[width=0.465\textwidth]{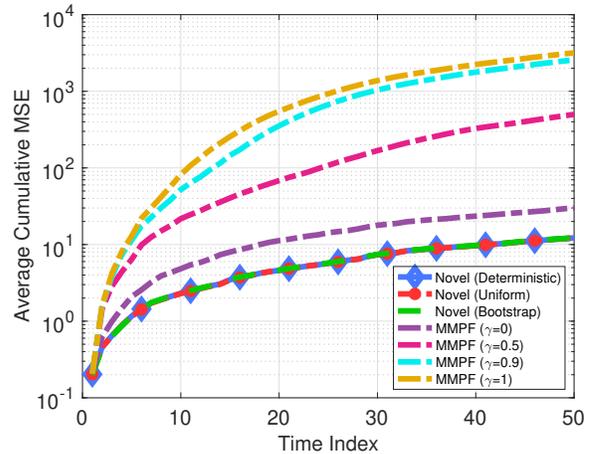}
\caption{Cumulative sum of MSE (Markovian dynamics).}
\label{fig: Markovian}
\end{figure}

Three different model index proposal distributions were used (deterministic, uniform, and bootstrap). For comparison, we also ran the multiple model particle filtering (MMPF) algorithm presented in \cite{liu2011instantaneous, liu2017robust}, where we drew 250 samples per model. Note that this algorithm considers a forgetting factor parameter $\gamma\in[0,1]$ that determines how much the observation history influences the model probabilities. The closer $\gamma$ is to 1, the more the observation history influences the model probabilities. We tested four different settings of this method, each corresponding to a different forgetting factor $\gamma\in\{0, 0.5, 0.9, 1\}$. The results are averaged over $500$ Monte Carlo runs and are summarized in Tables \ref{table: mse markovian} and \ref{table: model markovian}. We can see that the novel method, regardless of the choice of the model index proposal distribution, provides a smaller mean squared error (MSE) and more accurate model selection results. For reference, we also plot the average cumulative sum of the MSE in the state estimation in Fig. \ref{fig: Markovian}. 

\begin{table}[b!]
\centering
{\small
\begin{tabular}{|l|c|c|c|}
\hline
 & Average & Best & Worst \\
\hline
Novel (Deterministic) & 0.2443 & 0.0566 & 4.8527   \\
\hline
Novel (Uniform) & 0.2446 & 0.0573 & 4.7388 \\
\hline
Novel (Bootstrap) & 0.2462 & 0.0546 & 4.9030 \\
\hline
MMPF ($\gamma=0$) & 0.5986 & 0.0792 & 9.0508 \\
\hline
MMPF ($\gamma=0.5$) & 9.9912 & 0.2635 & 279.7545 \\
\hline
MMPF ($\gamma=0.9$) & 51.4122 & 1.5156 & 900.7603\\
\hline
MMPF ($\gamma=1$) & 63.5191 & 1.5136 & 1002.7609 \\
\hline
\end{tabular}
}
\caption{State estimation MSE (Markovian dynamics).}
\label{table: mse markovian}
\end{table}

\begin{table}[b!]
\centering
{\small
\begin{tabular}{|l|c|c|c|}
\hline
 & Average & Best & Worst \\
\hline
Novel (Deterministic) & 0.9407 & 1 & 0.5000 \\
\hline
Novel (Uniform) & 0.9402 & 1 & 0.5000 \\
\hline
Novel (Bootstrap) &  0.9419 & 1 & 0.5000 \\
\hline
MMPF ($\gamma=0$) & 0.8437 & 1 & 0.5200\\
\hline
MMPF ($\gamma=0.5$) & 0.5089 & 0.9200 & 0.1200 \\
\hline
MMPF ($\gamma=0.9$) & 0.2348 & 0.9200 & 0 \\
\hline
MMPF ($\gamma=1$) & 0.2180 & 0.9200 & 0 \\
\hline
\end{tabular}
}
\caption{Model selection accuracy (Markovian dynamics).}
\label{table: model markovian}
\end{table}

\begin{table}[h!]
\centering
{\small
\begin{tabular}{|l|c|c|c|}
\hline
 & Average & Best & Worst \\
\hline
Novel (Deterministic) & 0.4112 & 0.0663 & 2.3021 \\
\hline
Novel (Uniform) & 0.4111 & 0.0672  & 2.2512 \\
\hline
Novel (Bootstrap) & 0.4116 & 0.0644 & 2.3921  \\
\hline
MMPF ($\gamma=0$) & 0.4995 & 0.0734 &  2.6993  \\
\hline
MMPF ($\gamma=0.5$) & 4.4573 & 0.3854 & 28.0283 \\
\hline
MMPF ($\gamma=0.9$) & 8.6751 & 1.9418 & 32.9856 \\
\hline
MMPF ($\gamma=1$) & 11.1040 & 2.0316 & 43.6506  \\
\hline
\end{tabular}
}
\caption{State estimation MSE (P\'{o}lya urn dynamics).}
\label{table: mse polya}
\end{table}

\begin{table}[h!]
\centering
{\small
\begin{tabular}{|l|c|c|c|}
\hline
 & Average & Best & Worst \\
\hline
Novel (Deterministic) & 0.9003  & 1 & 0.6800 \\
\hline
Novel (Uniform) & 0.8996 & 1 & 0.6600 \\
\hline
Novel (Bootstrap) & 0.8996 &  1& 0.6800 \\
\hline
MMPF ($\gamma=0$) & 0.8526 & 1 &  0.6200 \\
\hline
MMPF ($\gamma=0.5$) & 0.2867  &  0.5600 & 0.0600 \\
\hline
MMPF ($\gamma=0.9$) & 0.1690 & 0.4200 & 0 \\
\hline
MMPF ($\gamma=1$) & 0.1571 & 0.4800 & 0\\
\hline
\end{tabular}
}
\caption{Model selection accuracy (P\'{o}lya urn dynamics).}
\label{table: model polya}
\end{table}

\begin{figure}[t]
\centering
 \includegraphics[width=0.465\textwidth]{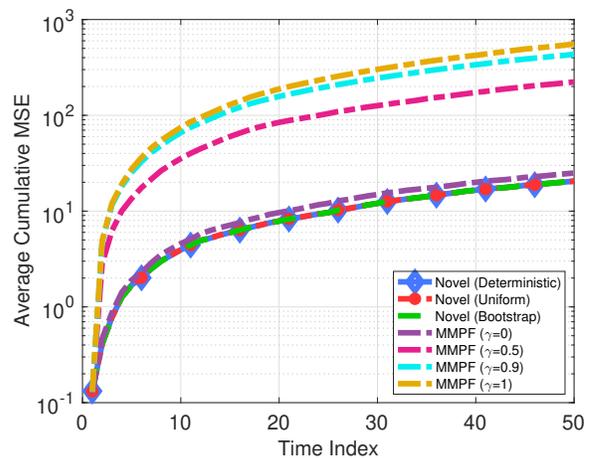}
\caption{Cumulative sum of MSE (P\'{o}lya urn dynamics).}
\label{fig: Polya urn}
\end{figure}

Next, we conducted the proposed method on the P\'{o}lya urn process. { The initial counts of the eight models were a random permutation of the integers from 1 to 8.} The parameter settings for RSPF and MMPF were the same as above. Table \ref{table: mse polya} and \ref{table: model polya} show the results, which are averaged over 500 Monte Carlo simulations. { Again, we can see that the novel method outperforms MMPF with the settings $\gamma=0.5, 0.9$ and $1$ by far, and slightly outperforms the MMPF with a forgetting factor of 0 in terms of model selection accuracy, and in terms of state estimation MSE. } The {average} cumulative sum of the MSE for the state estimation is shown in Fig. \ref{fig: Polya urn}.

\section{Conclusions}
\label{sec: conclusions}
In this paper, we introduced a novel particle filtering algorithm for regime switching systems. The proposed method allows for the treatment of stochastic filtering problems under model uncertainty, where the model can change from one time instant to the next. Moreover, our algorithm does not have any restrictions on the regime switching dynamics and can work for systems that are not Markovian switching systems. We validated our method on two synthetic data experiments, where in the first experiment we considered a Markovian switching system and in the second experiment we considered a system where regimes changed according to a P\'{o}lya urn process.  

\bibliographystyle{IEEEbib}
\bibliography{refs}

\end{document}